\documentclass[twoside,twocolumn,9pt]{article}
\usepackage[super,sort&compress,comma]{natbib}
\usepackage[left=1.5cm, right=1.5cm, top=1.785cm, bottom=2.0cm]{geometry}
\usepackage{balance}
\usepackage{times,mathptmx}
\usepackage{sectsty}
\usepackage{graphicx}
\usepackage{lastpage}
\usepackage[format=plain,justification=justified,singlelinecheck=false,font={stretch=1.125,small,sf},labelfont=bf,labelsep=space]{caption}
\usepackage{float}
\usepackage{fancyhdr}
\usepackage{fnpos}
\usepackage[english]{babel}
\usepackage{array}
\usepackage{droidsans}
\usepackage{charter}
\usepackage[T1]{fontenc}
\usepackage[usenames,dvipsnames]{xcolor}
\usepackage{setspace}
\usepackage[compact]{titlesec}
\usepackage{hyperref}
\usepackage{graphicx, subfigure} 
\usepackage{amsmath,amsfonts}

\definecolor{cream}{RGB}{222,217,201}

\begin{document}

\pagestyle{fancy}
\thispagestyle{plain}
\fancypagestyle{plain}{

\fancyhead[C]{\includegraphics[width=18.5cm]{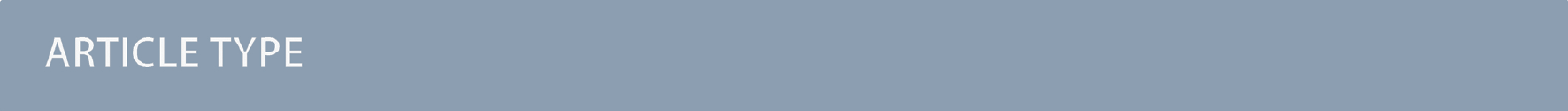}}
\fancyhead[L]{\hspace{0cm}\vspace{1.5cm}\includegraphics[height=30pt]{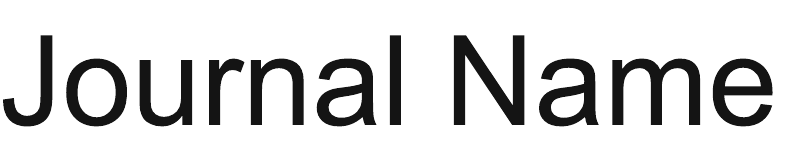}}
\fancyhead[R]{\hspace{0cm}\vspace{1.7cm}\includegraphics[height=55pt]{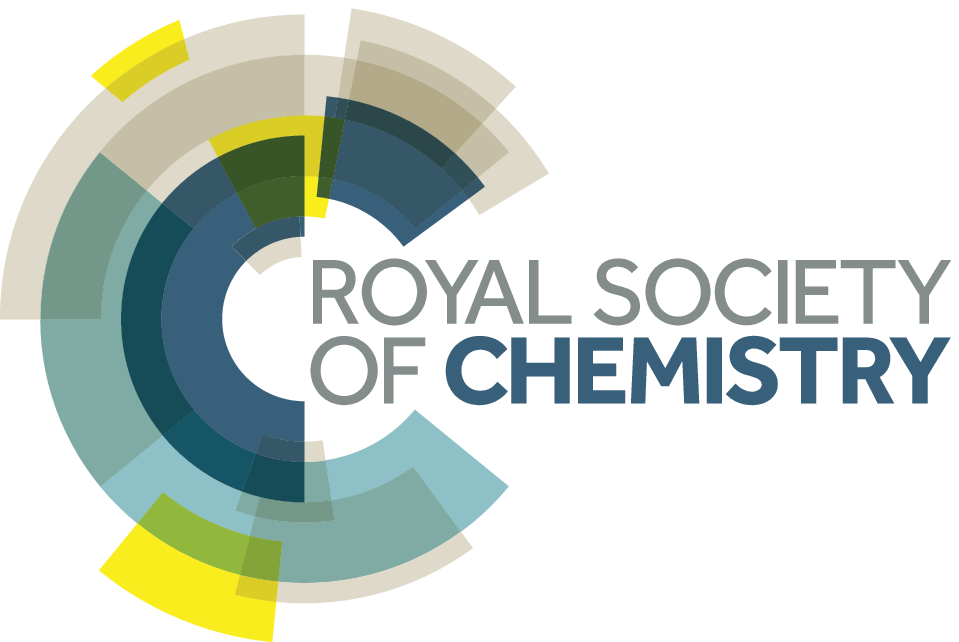}}
\renewcommand{\headrulewidth}{0pt}
}

\makeFNbottom
\makeatletter
\renewcommand\LARGE{\@setfontsize\LARGE{15pt}{17}}
\renewcommand\Large{\@setfontsize\Large{12pt}{14}}
\renewcommand\large{\@setfontsize\large{10pt}{12}}
\renewcommand\footnotesize{\@setfontsize\footnotesize{7pt}{10}}
\makeatother

\renewcommand{\thefootnote}{\fnsymbol{footnote}}
\renewcommand\footnoterule{\vspace*{1pt}%
\color{cream}\hrule width 3.5in height 0.4pt \color{black}\vspace*{5pt}}
\setcounter{secnumdepth}{5}

\makeatletter
\renewcommand\@biblabel[1]{#1}
\renewcommand\@makefntext[1]%
{\noindent\makebox[0pt][r]{\@thefnmark\,}#1}
\makeatother
\renewcommand{\figurename}{\small{Fig.}~}
\sectionfont{\sffamily\Large}
\subsectionfont{\normalsize}
\subsubsectionfont{\bf}
\setstretch{1.125} 
\setlength{\skip\footins}{0.8cm}
\setlength{\footnotesep}{0.25cm}
\setlength{\jot}{10pt}
\titlespacing*{\section}{0pt}{4pt}{4pt}
\titlespacing*{\subsection}{0pt}{15pt}{1pt}

\setlength{\headheight}{110.56966pt}

\fancyfoot{}
\fancyfoot[LO,RE]{\vspace{-7.1pt}\includegraphics[height=9pt]{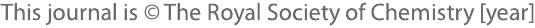}}
\fancyfoot[CO]{\vspace{-7.1pt}\hspace{13.2cm}\includegraphics{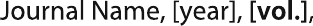}}
\fancyfoot[CE]{\vspace{-7.2pt}\hspace{-14.2cm}\includegraphics{head_foot/RF}}
\fancyfoot[RO]{\footnotesize{\sffamily{1--\pageref{LastPage} ~\textbar  \hspace{2pt}\thepage}}}
\fancyfoot[LE]{\footnotesize{\sffamily{\thepage~\textbar\hspace{3.45cm} 1--\pageref{LastPage}}}}
\fancyhead{}
\renewcommand{\headrulewidth}{0pt}
\renewcommand{\footrulewidth}{0pt}
\setlength{\arrayrulewidth}{1pt}
\setlength{\columnsep}{6.5mm}
\setlength\bibsep{1pt}

\makeatletter
\newlength{\figrulesep}
\setlength{\figrulesep}{0.5\textfloatsep}

\newcommand{\topfigrule}{\vspace*{-1pt}%
\noindent{\color{cream}\rule[-\figrulesep]{\columnwidth}{1.5pt}} }

\newcommand{\botfigrule}{\vspace*{-2pt}%
\noindent{\color{cream}\rule[\figrulesep]{\columnwidth}{1.5pt}} }

\newcommand{\dblfigrule}{\vspace*{-1pt}%
\noindent{\color{cream}\rule[-\figrulesep]{\textwidth}{1.5pt}} }

\makeatother

\twocolumn[
  \begin{@twocolumnfalse}
\vspace{3cm}
\sffamily
\begin{tabular}{m{4.5cm} p{13.5cm} }

\includegraphics{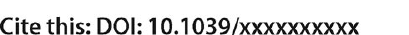} & \noindent\LARGE{\textbf{Self similarity of liquid droplet coalescence in a quasi-2D free-standing liquid-crystal film}} \\ 
\vspace{0.3cm} & \vspace{0.3cm} \\

 & \noindent\large{Christoph Klopp,$^{\ast}$\textit{$^{a}$} Torsten Trittel,\textit{$^{a}$} Ralf Stannarius$^{\dag}$\textit{$^{a}$}} \\

\includegraphics{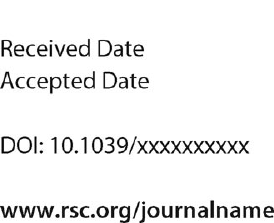} & \noindent\normalsize{

Coalescence of droplets is an ubiquitous phenomenon in chemical, physical and biological systems. The process of merging of liquid objects has been studied during the past years experimentally and theoretically in different geometries. We introduce a unique system that allows a quasi two-dimensional description of the coalescence process, micrometer-sized flat droplets in freely suspended smectic liquid-crystal films. We find that the bridge connecting the droplets grows linearly in time during the initial stage of coalescence, both with respect to its height and lateral width. We also  verify self-similar dynamics of the bridge during the first stage of coalescence. We compare our results with a model based on the thin sheet equations.
}\\
\end{tabular}

 \end{@twocolumnfalse} \vspace{0.6cm}
 ]
\renewcommand*\rmdefault{bch}\normalfont\upshape
\rmfamily
\section*{}
\vspace{-1cm}


\footnotetext{\textit{$^{a}$~Institute of Physics, Otto von Guericke University, Department of Nonlinear Phenomena, D-39106 Magdeburg, Universit\"ats\-platz 2, Germany.}}

\footnotetext{$\ast$~christoph.klopp@ovgu.de, $\dag$~ralf.stannarius@ovgu.de}


\section{\label{sec:level1}Introduction}
The motion of fluid structures within other fluids in restricted geometries as well as the flow surrounding them are interesting phenomena in physical, chemical and biological systems. Experimental studies of merging fluid objects started at the end of 19$^{\rm th}$ century, with pioneering work on collisions of liquid jets by Rayleigh \cite{Rayleigh1879,Rayleigh1879b,Rayleigh1883}, on interactions of touching soap bubbles \cite{Boys1888,Kaiser1894}, and others. In some cases, droplets in contact with surfaces of the same liquid avoid merging (see, e.~g. \cite{Mahajan1929,Derjaguin1993,Gilet2012}), but in most situations, droplets coalesce after contact, reducing their surface area by forming one single, larger drop.
The merging of liquid objects such as droplets, fluid cylinders or flat disks has important practical relevance
(see, e.~g. \cite{Kamp2017} and Refs. therein). One can find coalescence, for example, during rain drop formation \cite{eggers3} but also in the process of merging powder into a homogeneous material by heating (sintering) \cite{Skorokhod1996}. Industrial applications, such as ink-jet printing, coating processes and the stabilization of emulsions require a better and detailed understanding of the merging dynamics of liquid droplets. Several investigations of three-dimensional (3D) and 2D coalescence of liquid objects and their dynamics, published particularly during the last two decades, were devoted to  the description of these processes 
\cite{Hopper1984,Hopper1993,Hopper1993b,Wu2004,Yao2005,paulsen1,paulsen2,Paulsen2013,Bradley1978,burton1,Sprittles2012,Sprittles2014,Zhang2015,Irshad2018,shuravin1,Delabre2008,Delabre2010,WookLee2012,Kapur2007,Eddi2013,Sui2013,Zhang2015b,Moghtadernejad2015,Moghtadernejad2016,Somwanshi2018,Bruning2018,Pawar2019}. 

A model that analytically describes the coalescence of infinitely extended coaxial cylinders has been developed by Hopper \cite{Hopper1984,Hopper1993,Hopper1993b}. This model has been applied, with qualitative but no quantitative agreement, to describe the 
merging of so-called islands, flat circular disks of surplus material, in the plane of freely suspended 
smectic liquid-crystal films\cite{shuravin1,NguyenPhD,Stannarius2017}: 
A similar geometry is that of flat nematic islands floating on an immiscible liquid 
\cite{Delabre2008,Delabre2010}. In that case, the flow of the nematic material during coalescence is coupled to flow of the substrate liquid, and the authors observe a transition from an initial dynamics driven by surface dissipation to a later stage with volume dissipation, which changes the exponent of the observed scaling laws.

Another frequently encountered geometry is that of coalescing spherical or suspended hemispherical droplets in 3D \cite{Bradley1978,burton1,Wu2004,Yao2005,paulsen1,paulsen2,Paulsen2013,Sprittles2012,Sprittles2014,Zhang2015}.
In few of these experiments, spherical droplets are freely floating in a carrier liquid \cite{Bradley1978,burton1,Zhang2015}, while in most cases (e. g.~\cite{Wu2004,Yao2005,paulsen1,paulsen2,Paulsen2013} sphere-cap shaped droplets are suspended on nozzles. 
The initial phases of coalescence are comparable, but the final stage of the latter is a catenoid connecting the supporting nozzles.

Droplets attached to flat solid substrates show a qualitatively different dynamics of merging, mainly because the flow field is qualitatively 
different from that in the coalescence of free droplets,
owing to no-slip conditions for the flow field at the solid surface \cite{Kapur2007,WookLee2012,Eddi2013,Sui2013,Zhang2015b,Moghtadernejad2015,Moghtadernejad2016,Somwanshi2018,Bruning2018,Pawar2019}. The wetting characteristics of the fluid on the substrate is relevant and the contact angle becomes an important parameter.

In addition to solid substrates, there are also experiments of droplets coalescing on liquid substrates \cite{Mitra2015,Zhang2019,Hack2019}, including
merging of droplets with a planar surface of the same liquid \cite{Zhang2019}.
 
In the present paper, we analyze the coalescence dynamics of liquid droplets in a quasi two-dimensional geometry. Flat, lens-shaped droplets are embedded in a nanometer thin free-standing liquid-crystal film. We observe the micrometer sized droplets using high-speed interferometric microscopy. This geometry has several advantages over previous experiments: First, there is no substrate or subphase, the film is embedded
in air on both sides. Thus, one can assume that the flow profile during coalescence is practically two-dimensional, and no velocity gradients normal to the film plane exist. The components of the velocity vectors are limited to the film plane. The vertical component can be neglected in the dynamic equations. The evolution of the lens height is governed by volume conservation. Second, the observation in reflected monochromatic light allows to record the contour of the merging droplets on the film as well as the simultaneous detection of their height profile from interference fringes. Thus, one can extract the complete three-dimensional profile of the merging droplets. By changing the temperature near the clearing point of the smectic mesophase, one can control the 
contact angle of the droplets with the film without changing any other parameters in the coalescence process. All these features make the present system uniquely suitable to test and confirm models describing the coalescence of such liquid lenses. 

Even though this system is at first glance related to the coalescing smectic islands studied by Shuravin et al. \cite{shuravin1} and Nguyen et al.~\cite{NguyenPhD,Stannarius2017}, there is a fundamental difference between both phenomena: The islands are two-dimensional flat disks of constant height whose coalescence is driven by the line tension of
the dislocations surrounding them.
The island material is in the same mesophase as the surrounding film. The smectic island coalescence was found to behave very similar to the predictions by Hopper for infinitely long coalescing parallel cylinders.

In our experiment, the droplets are formed from a different phase of the material, the driving force
arises from the reduction of the surface energy of the droplets, while the line tension at the contact to the surrounding film is negligible. Additionally, the height profile of the liquid lenses and the 3D character of the process cannot be neglected. Both the top views and the height profiles of the liquid bridges that connect the droplets during merging can be observed simultaneously, and we are able to construct fully 3D images at all time steps. The experimental results are compared to an 
analytical model by Hack et al. \cite{Hack2019}. This model is based on the lubrication flow in the cross-section normal to the film plane, along the axis of initial connection between the droplets. It predicts the dynamics and the self-similarity of the liquid bridge. 

\section{\label{sec:level2}Experimental setup}

Our experiments are performed with free-standing films in the smectic A phase. Such LC films can be prepared with areas up to several square centimeters and thickness from only $\approx 10$~nm up to {several micrometers}. Due to the layered structure of the smectic A phase, these films are homogeneous in thickness and extremely robust.  In fact, our films can be considered as quasi-2D. Liquid inclusions in the form of flat lenses embedded in the film plane are prepared with a procedure described below. 

\begin{figure}[ht!]
\centering
\includegraphics[width=0.8\columnwidth]
{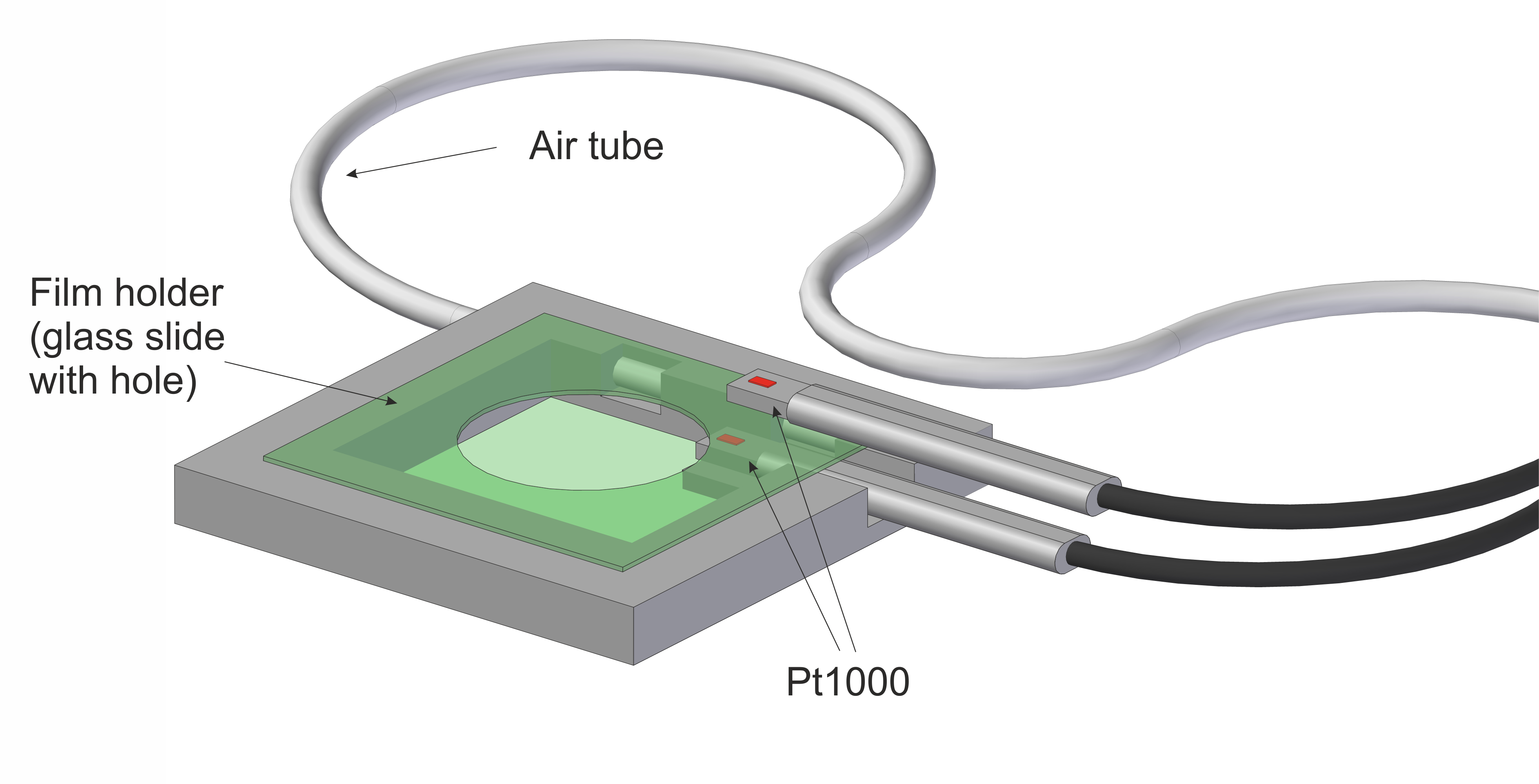}
	\caption{Sketch of the pressure chamber that holds the smectic film. Air pressure can be adjusted in the chamber to bend the film upwards of downwards. Two PT 1000 temperature sensors measure the temperature above and below the film. The hole in the support glass has a diameter of 10 mm. This chamber is inserted in a Linkam THMS 600 heating stage
}
	\label{fig:setup1}
\end{figure}

The smectic freely suspended films are drawn across a circular hole of 10 mm diameter in the top plate of a 4 mm high pressure chamber (Fig.~\ref{fig:setup1}), placed on a controllable hot stage (Linkam THMS 600). After a film is drawn, the air below the film in the airtight chamber can be partially evacuated with a microsyringe. Thereby, the film will slightly bend down until the film curvature creates a Laplace pressure compensating the underpressure in the chamber. 
 
The complete setup is placed under a polarizing microscope (ZEISS Axioscope 40) where we illuminate the film with a mercury lamp and a green filter for monochromatic light. A Phantom VEO 710L high-speed camera is employed to observe the coalescence with a frame rate of 24 000~fps with a typical frame size of 500 x 500 pixels. With the color filter of wavelength $\lambda$=546~nm we are able to observe interference between light reflected at the top and bottom surfaces of the films or flat droplets. This allows us to determine local thicknesses of the structures with an
accuracy of about 90~nm.

\begin{figure}[ht!]
	\includegraphics[width=1\columnwidth]{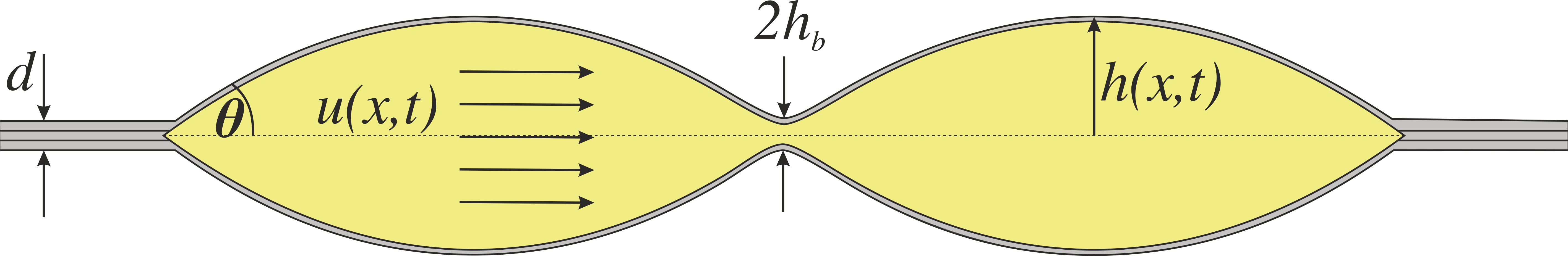}
	\caption{Sketch of two droplets at the beginning of coalescence, with bridge height $h_b(t)$ (half the bridge thickness) and the height profile $h(x,t)$ (half the droplet thickness) in the symmetric midplane. The isotropic droplets are embedded in the smectic A film and their surfaces are covered  by a few smectic layers. The vertical dimension is greatly exaggerated in the sketch}
	\label{fig:sidesketch}
\end{figure}

The liquid crystal material used is a mixture of 80~\% 5-heptyl-2-[4-(4-methylhexyloxy)-phenyl]-pyrimidine and 20~\% 4-(5-octyl-pyrimidine-2-yl)-benzoic acid decyl ester (Displaytech mixture MX 12160, shear viscosity $\eta = 0.014$ Pa$\cdot$s close to the transition temperature to the isotropic phase for shear flow in the film plane~\cite{klopp2}, {surface tension} $\gamma_{\rm sm} = 0.024$~N/m).
This thermotropic mixture has a bulk transition from the smectic A (SmA) to the isotropic phase at a temperature of 54$^\circ$C (clearing point). When this bulk transition temperature is reached, the inner layers of freely suspended films start to melt while the outer layers remain in the SmA phase~\cite{Pankratz1999,Mirantsev1995,Kranjc1996,Stoebe1994}. The isotropic material then collects in the form of liquid droplets embedded in the SmA environment (Figure \ref{fig:sidesketch}). At constant temperature, an equilibrium between the molten isotropic material and the remaining film is reached and the droplets coexist with the SmA film.

The initial droplet diameters $D_0$ are in the range from 5~$\mu$m to 100~$\mu$m and the droplets have initial heights $H_0$ between 0.2~$\mu$m and 2.3~$\mu$m. The upper and lower parts of the droplets are mirror-symmetric, they represent sphere caps in equilibrium. Note that $H_0$, $h_b$ and $h$ 
refer to the height of the surface respective to the film mid-plane, i. e. to half the local thicknesses. Knowing the geometrical parameters, one can calculate the contact angle $\theta$ between the film and the droplet surface:
\begin{equation}\label{eq:1}
\theta = 2\cdot \arctan(2H_0/D_0).
\end{equation}

The droplets are prepared on flat films by heating the material to slightly above the bulk transition temperature. Then, the film is slightly bent down by sucking some air out of the chamber. The radius of curvature of the film is well in the centimeter range so that the curvature can be neglected on the scale of the droplets. However, the droplets slide down towards the lowest region of the film. 
 By changing the curvature of the film with small pressure changes we can adjust the effective gravitational force in order to control the motion of the droplets towards each other. They touch each other and usually remain in this intermediate contact state for a few moments. After they have overcome a certain barrier, their coalescence starts. This is similar to the observations of Shuravin et al. \cite{shuravin1}
 who observed smectic islands coalescing in similar films. 
 
 \begin{figure}[ht!]
    \includegraphics[width=1\linewidth]{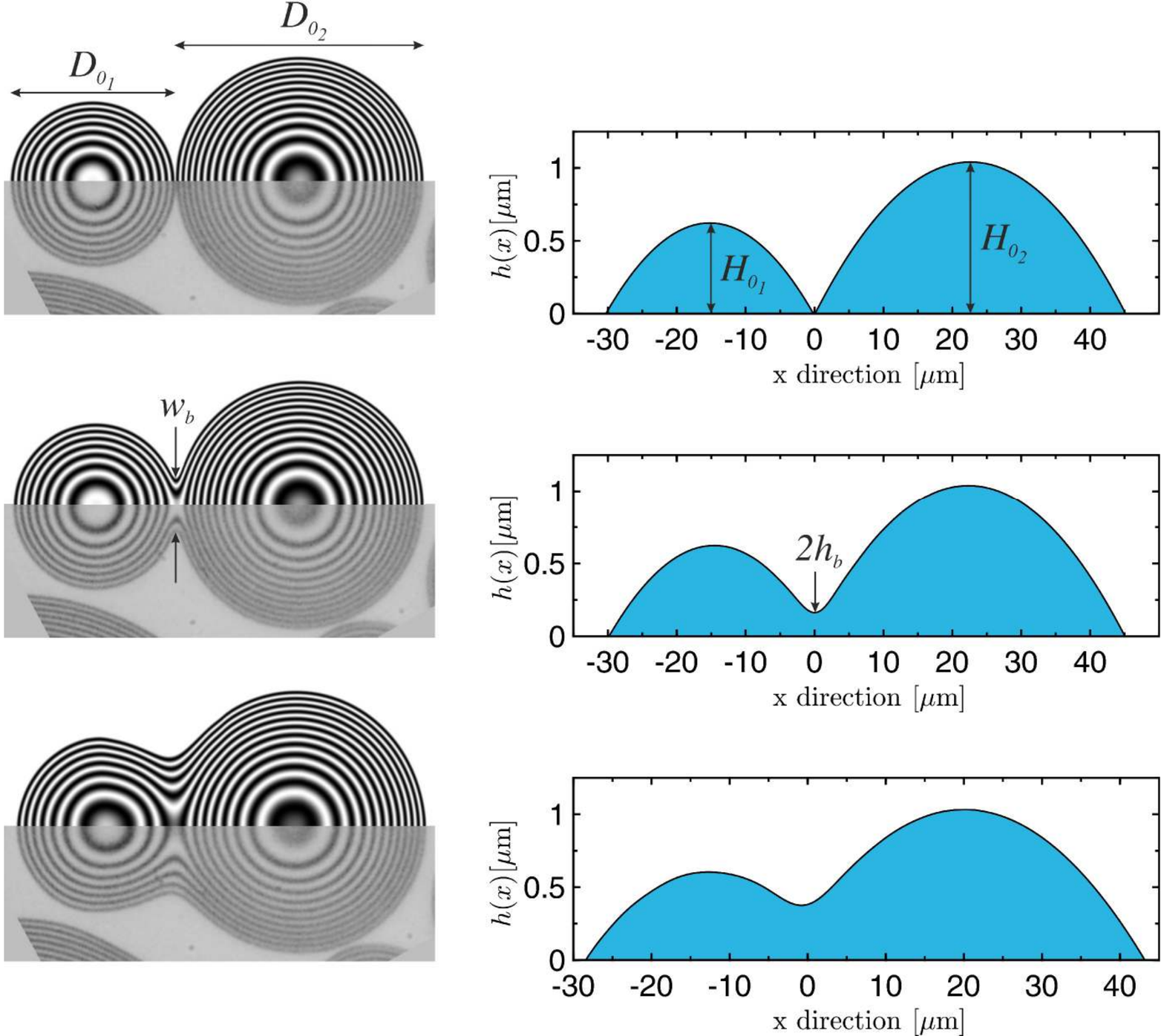}
    \caption{Top view of two merging droplets with interference fringes in monochromatic light ($D_{0{1}}=30.3~\mu$m, $D_{0{2}}=45.1~\mu$m, $H_{0{1}}=0.66~\mu$m, $H_{0{2}}=0.99~\mu$m). {Indices $1,2$ refer to the two droplets.} The snapshots were taken at times 0.0~ms, 0.5~ms and 1.0~ms after the start of the coalescence (from top to bottom). In each interference pattern, the bottom half is the experimentally observed picture, and the top half represents the calculated interference pattern using the fitted height profile. The graphs on the right hand side show the calculated height profiles in the symmetry axis through the centers of the droplets.} 
	\label{fig:rings}
\end{figure}

Interference rings of the droplets under monochromatic illumination provide the local droplet heights during the coalescence process. Based on these fringes, we can reconstruct the complete height profiles (Figure \ref{fig:rings}).

\section{\label{sec:level3}Results}
\begin{figure}[ht!]
	\centering
	\includegraphics[width=0.8\linewidth]{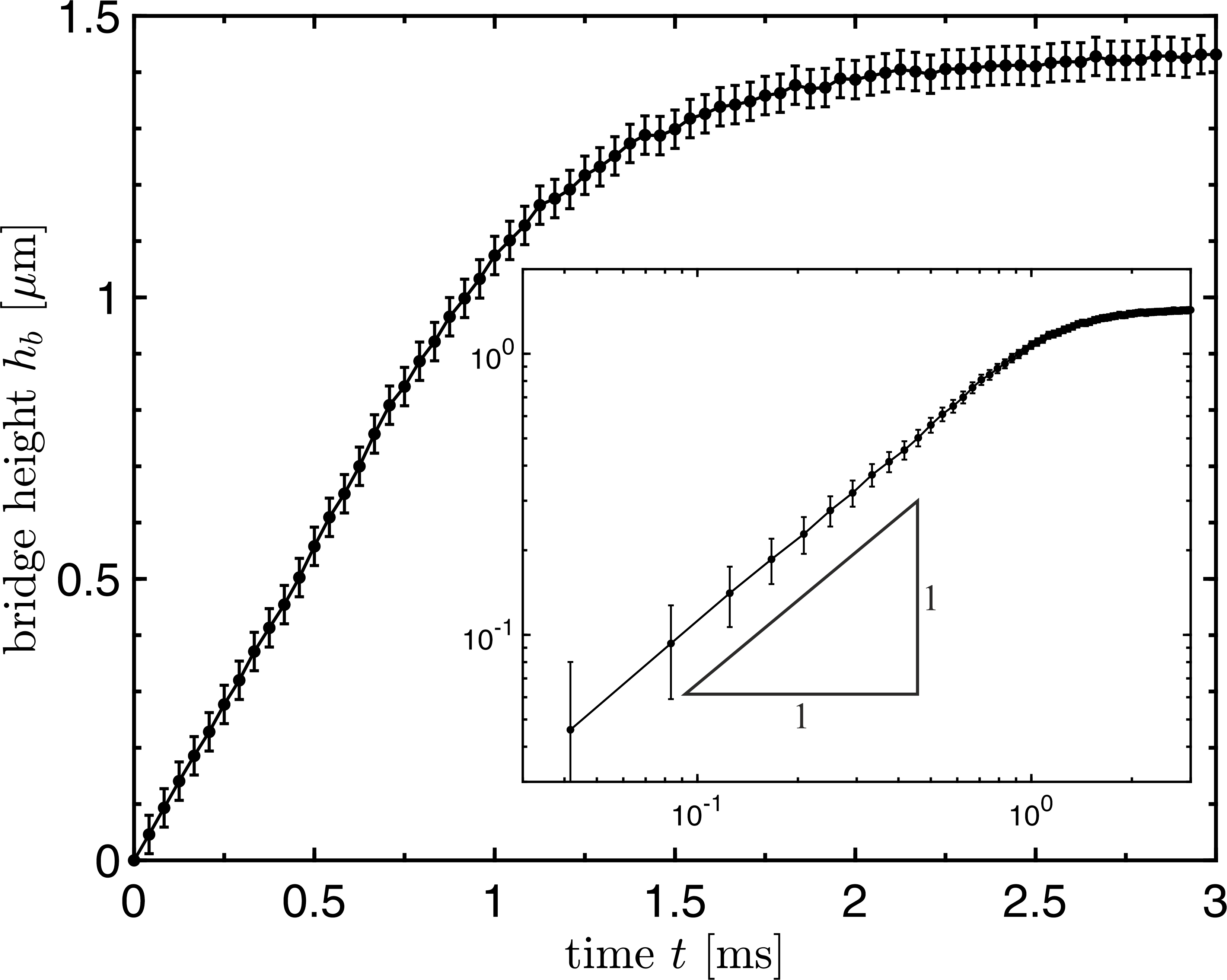}
	\caption{Measured bridge height $h_b$ during the merging process of two nearly equal-sized droplets ($D_{0{1}}=35.5~\mu$m, $D_{0{2}}=34.7~\mu$m, $H_{0{1}}=1.13~\mu$m, $H_{0{2}}=1.10~\mu$m, $\theta=7.3 ^\circ$). 
	}
	\label{fig:bridge_height}
\end{figure}

\begin{figure}[ht!]
	\centering
	\includegraphics[width=0.8\linewidth]{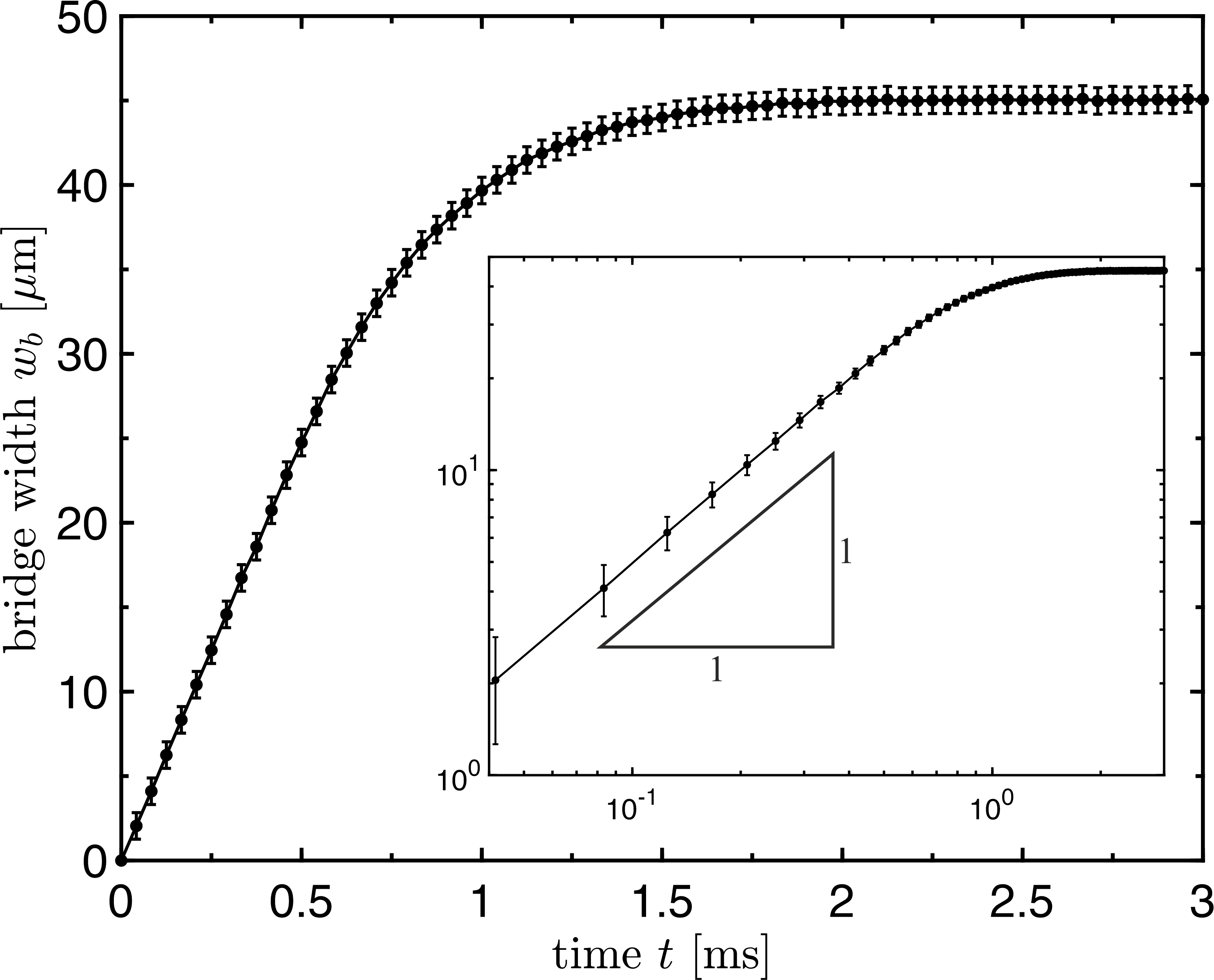}
	\caption{Measured bridge width $w_b$ during the merging process of two nearly equal-sized droplets ($D_{0{1}}=35.5~\mu$m, $D_{0{2}}=34.7~\mu$m, 
	$H_{0{1}}=1.13~\mu$m, 
	$H_{0{2}}=1.10~\mu$m,
	$\theta=7.3 ^\circ$).
	}
	
	\label{fig:bridge_width}
\end{figure}

Our experiments show that the connecting bridge grows linearly in time, both parallel to the film plane (width $w_b$) and perpendicular to it (height $2h_b$) during the first milliseconds. Figures \ref{fig:bridge_height} and \ref{fig:bridge_width} show these characteristics for a typical droplet pair. The log-log plots (insets of Fig.~\ref{fig:bridge_height} and Fig. \ref{fig:bridge_width}) confirm that linear trend. Hack et al. \cite{Hack2019} distinguished two regimes in their model: a nonlinear inertia dominated regime with small viscosities where they find a growth of the bridge height with $h_b\propto t^{2/3}$, and a linear regime where $h_b\propto t$ for high viscosities. They used the thin sheet equations, which implies two assumptions: the fluid flow during the whole coalescence process is dominated by flow in the film plane and velocity gradients normal to the plane can be neglected. 
the flow perpendicular to the film plane can be eliminated from the equations. Similar to drop coalescence on a substrate~\cite{Hernandez2012,Eddi2013}, the modelling of liquid lenses can be immensely simplified: The evolution of the bridge height is described using a two dimensional cross-section containing the connecting axis of the droplet centers (the $x$-axis in our coordinate system). Furthermore, Hack et al. assumed that the droplet merging is controlled and defined by the flow inside the droplets and that the flow in the subphase can be neglected. With these simplifications, the thin sheet equations read \cite{erneux}:
\begin{eqnarray}
h_t +(uh)_x=0 \label{eq:lu1}\\
\rho  (u_t +u u_x) = { \gamma h_{xxx}}+4{\eta} \frac{(h u_x)_x}{h}
\label{eq:lu2}\end{eqnarray}
where $h(x,t)$ defines the shape of the merging droplet depending on time and position and $u(x,t)$ is the velocity parallel to the free standing film. {Lower indices indicate spatial and temporal derivatives, where $x$ is taken along the direction connecting the two droplet centers, see Fig.~\ref{fig:rings}.  Plug flow is assumed} in first approximation. Within that equation that includes mass and momentum conservation, $\rho$ is the density, and $\gamma$ is the surface tension of the droplets with respect to the surrounding air. {This value is slightly larger than $\gamma_{\rm sm}$ because of the existence
of an additional interface tension at the droplet surface between
the smectic skin layers and the isotropic bulk liquid inside the
droplets.}

For the early shape evolution, Hack et al. \cite{Hack2019} solved these equations for the present geometry of two equally sized coalescing droplets introducing similarity solutions of the form
\begin{equation}\label{eq:scaling}
h(x,t)=kt^{\alpha}\mathcal{H}(\xi), ~~~ u(x,t)=\dfrac{{\alpha}k}{\theta}t^{\beta}\mathcal{U}(\xi), ~~~ \xi=\dfrac{\theta x}{kt^{\alpha}},
\end{equation}
where $\mathcal{H}$ and $\mathcal{U}$ are the self-similarity functions for the bridge height profile and the flow velocity in the droplets. The parameter $\xi$ is chosen such that $h(x,t)$ {reaches the contact angle $\theta$} far from the bridge.
For the viscous regime which is relevant here, they set $\rho =0$ {(neglect the inertial term)} and found $\alpha =1$, $\beta =0$, thus 
\begin{equation}
h(x,t)=kt\mathcal{H}(\xi), ~~~ u(x,t)=\dfrac{k}{\theta}\mathcal{U}(\xi), ~~~ \xi=\dfrac{\theta x}{kt}.
\end{equation}
As one result of the analysis, the model predicts a linear growth of the bridge height during the initial phase of coalescence. In the experiment, we observed this trend shown in Fig.~\ref{fig:bridge_height} over a time period of about 1 ms. The nonlinear deviations set in only when the bridge height has already reached more than 3/4 of the final droplet height.

An important parameter of that model is the dimensionless bridge velocity
\begin{equation}\label{eq:5}
K=\dfrac{4\eta k}{\gamma \theta^2}
\end{equation}
Here, $k={d h_b}/{d t}$ represents the dimensional growth velocity of the bridge height. From the numerical solution, Hack et al. \cite{Hack2019} found a value of $K\approx 2.21$ as the correct value of the shooting parameter that satisfies the boundary conditions.

Using Eq.~(\ref{eq:5}), we can describe the initial growth rate of the bridge height ${d h_b}/{d t}$ in dependence of the contact angle $\theta$,
\begin{equation}
\dfrac{d h_b}{d t}= \dfrac{ {K} \gamma}{4 \eta} \theta ^2.
\end{equation}
Note that this parameter depends in quadratic form on the contact angle. The latter is related to the ratios of interface tensions of the droplet surfaces and the surrounding smectic film by Young's equation. Our system allows a straightforward adjustment and measurement of $\theta$ and thus a direct 
test of Eq.~(\ref{eq:5}). In the system described in Ref. \cite{Hack2019}, where oil droplets on a water subphase were considered, the change of the contact angle requires chemical modifications of the fluids or addition of surfactants. In our smectic films, the contact angles are very sensitive to temperature changes near the phase transition \cite{schuering1}. When the temperature is varied by a few tenth of a degree, the contact angle can be changed substantially. Thus, the predicted square dependence of the dimensionless coalescence rate $K$ in Eq.~(\ref{eq:5}) can be verified experimentally. 
While the actual temperatures can be {\em adjusted} within a few dozen mK, the {\em measurement} of the exact film temperature with the same precision is not possible. Nevertheless, the widths of the interference fringes (Fig.~\ref{fig:sidesketch}) are accurate indicators of the contact angles, which we control by setting the film temperature~\cite{schuering1}. 

\begin{figure}[ht!]
	\centering
	\includegraphics[width=0.99\linewidth]{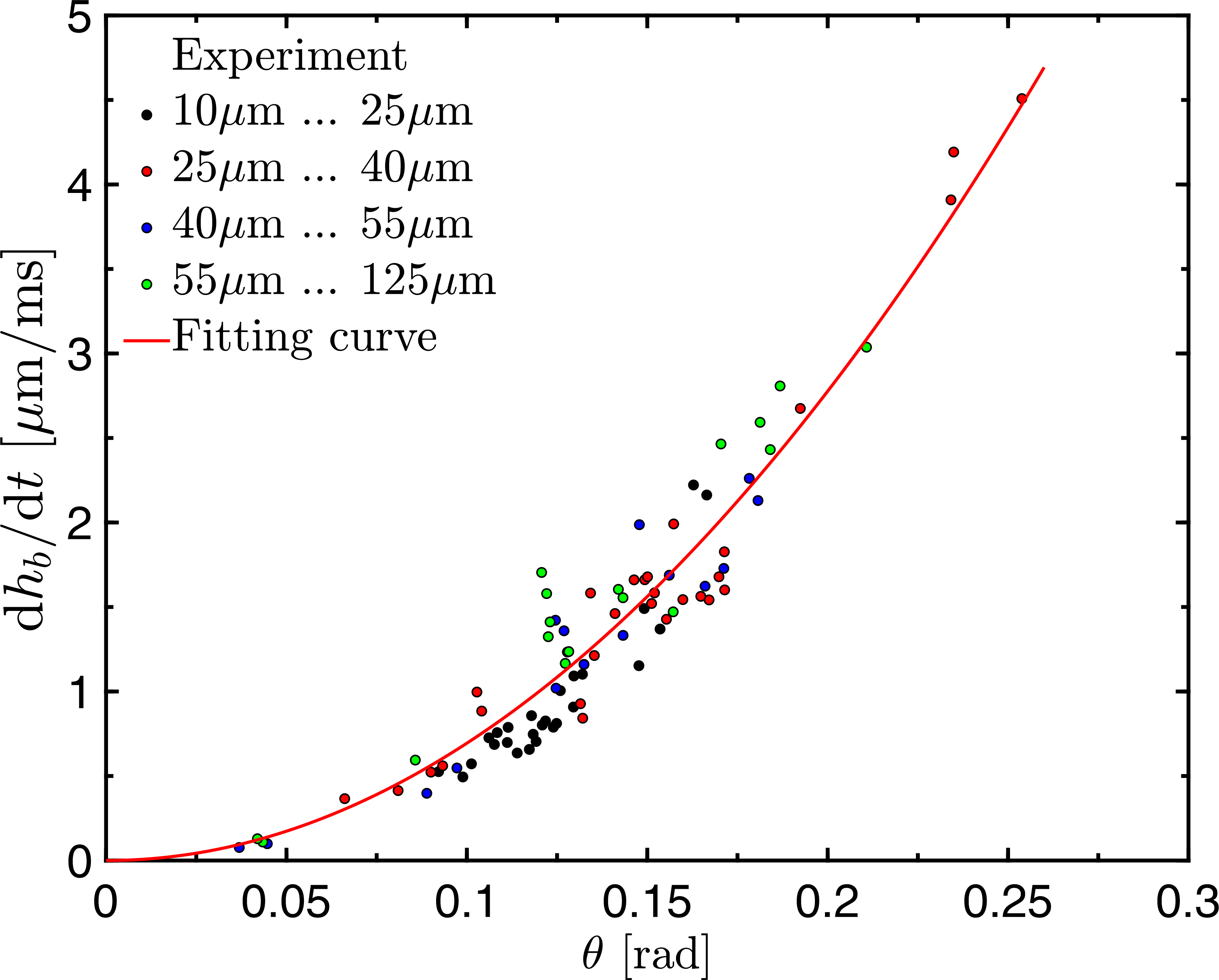}
	\caption{Growth rate of the bridge height $h_b$ in dependence of the contact angle $\theta$. The different colors in the experimental data visualize different diameter ranges of the merged (final) droplets. From a quadratic fit (solid red curve), we extract the parameter  ${K}\gamma/\eta=0.28$~m/s}
	\label{fig:slope_height}
\end{figure}

\begin{figure}[ht!]
	\centering
	\includegraphics[width=0.99\linewidth]{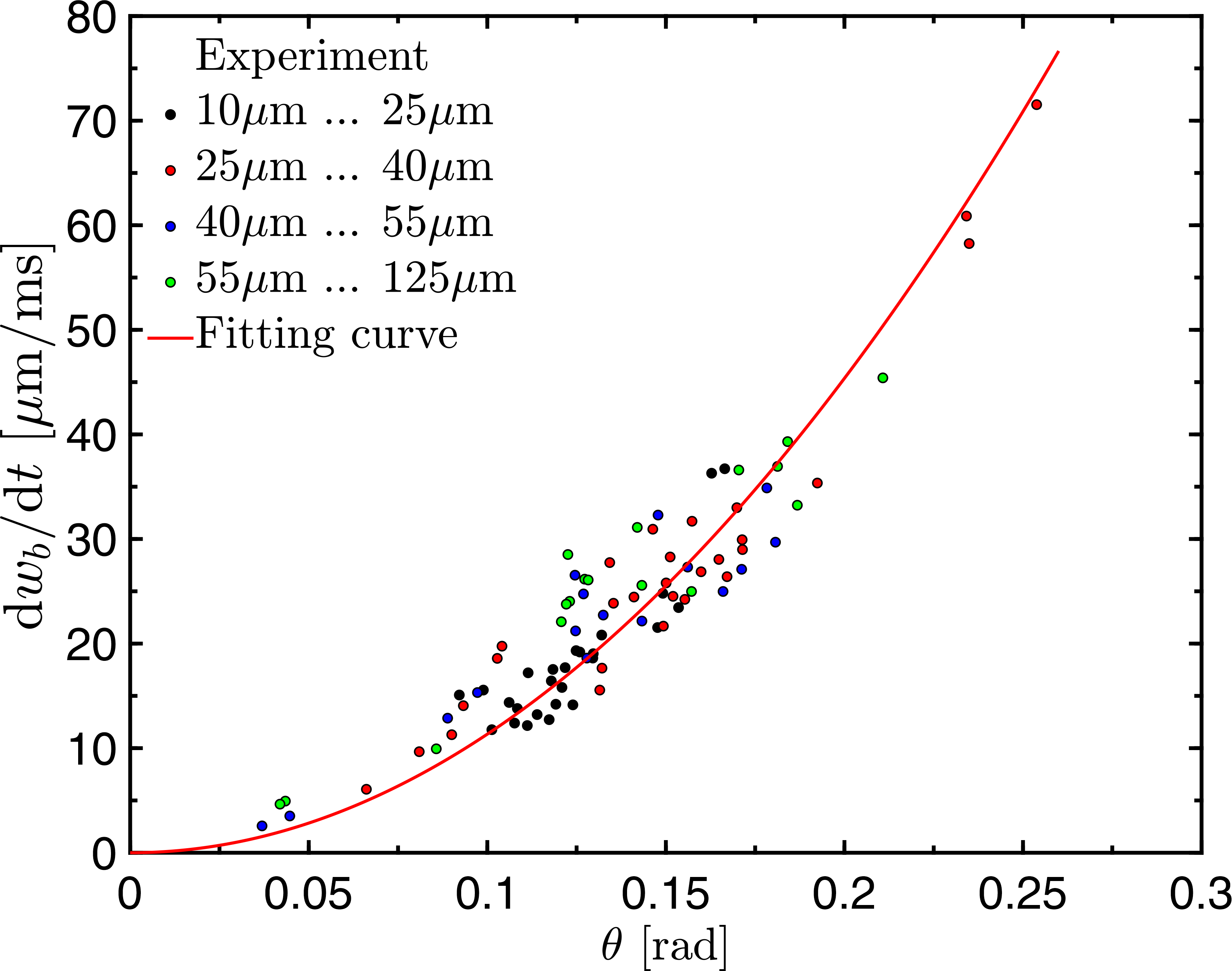}
	\caption{Growth rate of the bridge width {in dependence of the contact angle. The different colors in the experimental data visualize different diameter ranges of the merged (final) droplets. Since the profile of the bridge does not vary much during the coalescence, one can fit a quadratic curve to the width characteristics as well (solid red curve), the ratio of bridge height and width is roughly constant, between 0.045 and 0.06 (see Fig.~\ref{fig:contact})}}
	\label{fig:slope_width}
\end{figure}

Figures \ref{fig:slope_height} and \ref{fig:slope_width} show the measured growth rates of bridge heights and widths for different droplet pairs as functions of the contact angle. The experiments confirm the quadratic dependence of $dh_b/dt$ on $\theta$,
but there is a quantitative discrepancy with the 
model. If one considers the result $K = 2.21$ from Ref.~\cite{Hack2019} for the bridge height growth rate, then the ratio $\gamma/\eta$ would have to be 0.125 m/s. With the known surface tension of the liquid-crystal material, $\gamma = 0.024$ N/m and the shear viscosity {$\eta = 0.014$ Pa\,s}, one expects $\gamma/\eta\approx1.7$,
i. e. more than one order of magnitude larger. In fact, the smectic material obviously obeys the same scaling as predicted by the lubrication model, but it flows one order of magnitude slower than predicted. 

\begin{figure}[ht!]
	\centering
	\includegraphics[width=0.9\linewidth]{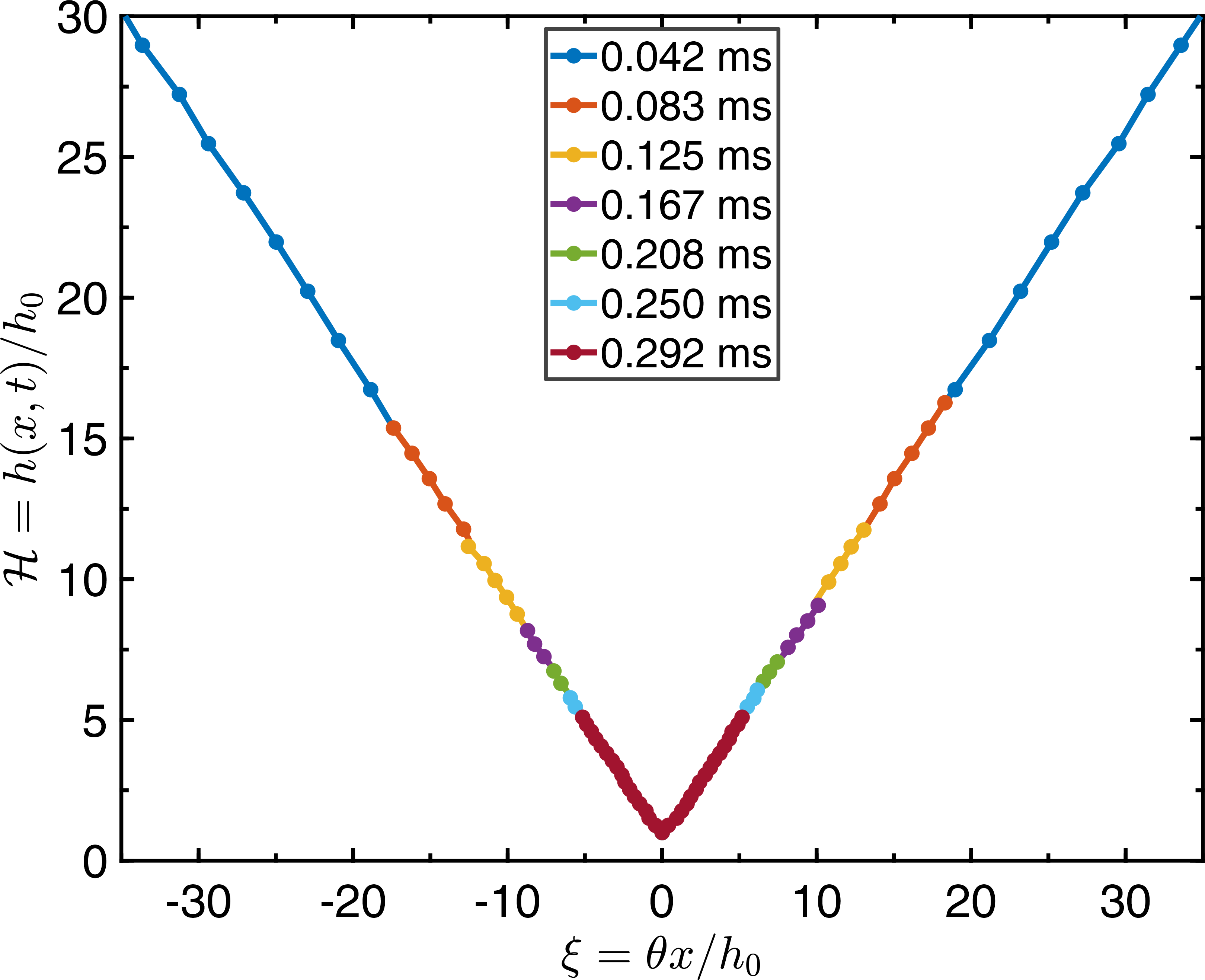}
	\caption{Scaled height profile at different instants at the position of the bridge center ($x$=0), all curves collapse to one single curve, which verifies the  self-similarity of the height profile ($D_{0_{1}} =65.5~\mu$m, $D_{0_{2}}=101.8~\mu$m, $H_{0_{1}}=1.41~\mu$m, $H_{0_{2}}=2.19~\mu$m), 
	from the center of the bridge we plotted the profile 
	12 micrometer in both directions}
	\label{fig:similar_height}
\end{figure}

Figure \ref{fig:similar_height} confirms that the scaling of the height profiles introduced in Eqs.~(\ref{eq:scaling}) describes the initial stages of coalescence excellently, even better than for the floating oil lenses in Ref.~\cite{Hack2019}.
 

An important difference to the experiments of Hack et al. \cite{Hack2019} is the free standing film in our system. In their experiments, they studied droplets floating on a liquid surface, and the bottom half of the lenses was completely embedded in the liquid subphase. Using thin films, our droplets are only covered by a few nanometers of smectic material, but otherwise surrounded by air only. Practically, these two situations are equivalent, except that in our system, the geometry is symmetric respective to the film plane. The surface tension of the subphase respective to air in the previous experiment plays the role of the smectic film tension. The latter does not explicitly enter the equations (\ref{eq:lu1},\ref{eq:lu2}), it is only implicitly included via Young's equation fixing the lenses' contact angle. The gain in energy by the shortening droplet contact line is much smaller than the reduction of the surface energy, i.e. the line tension along the droplet boundaries is negligible. This again highlights the substantial difference between coalescing droplets and coalescing islands in smectic films.

\begin{figure}[ht!]
	\centering
	\includegraphics[width=0.8 \linewidth]{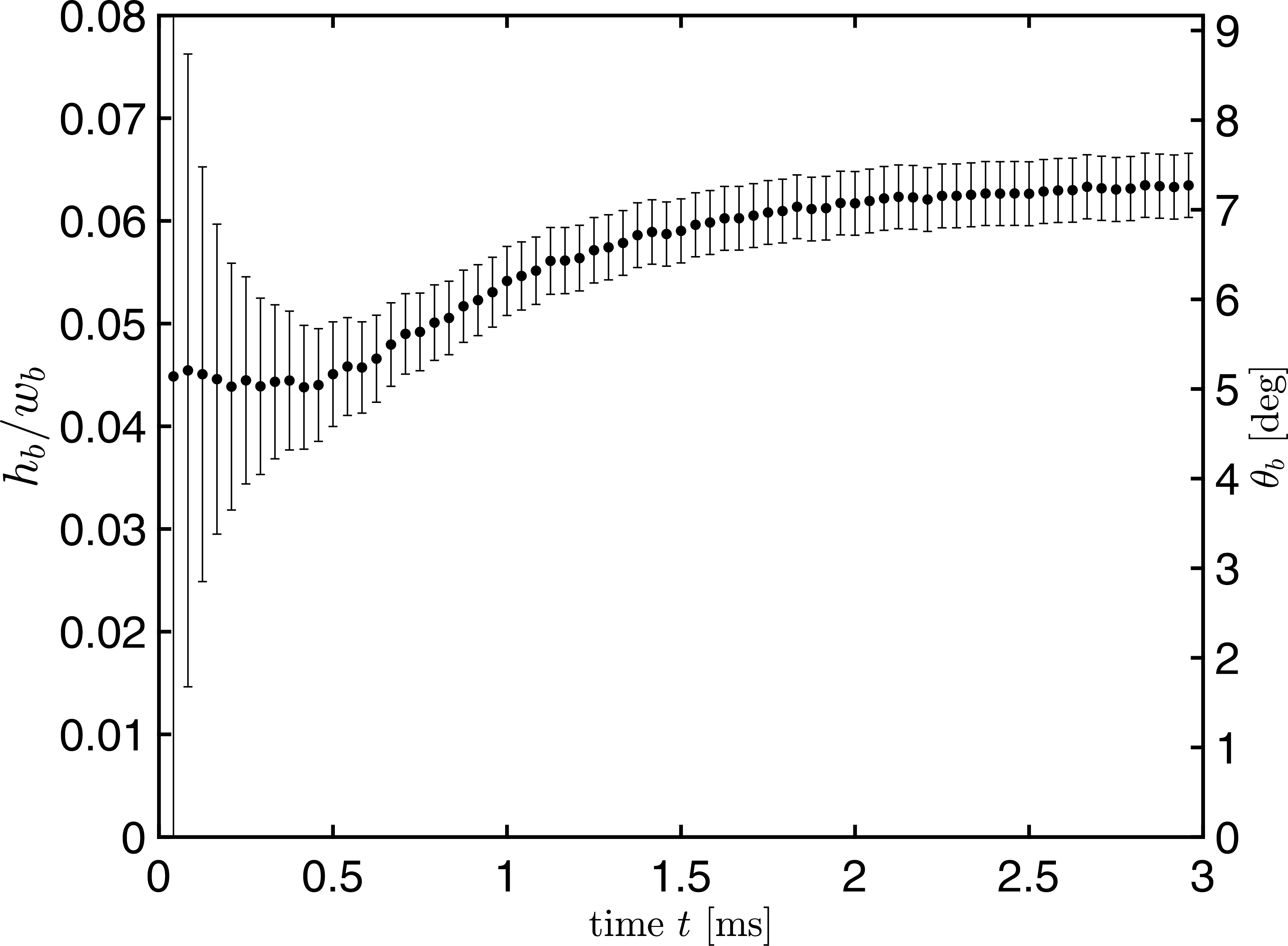}
	\caption{Ratio of bridge height and width, the right axis shows the contact angle $\theta_b$ determined from this ratio with Eq.~(\ref{eq:1})  during coalescence - it finally reaches the static contact angle}
	\label{fig:contact}
\end{figure}

The $\theta^2$ dependence of the growth rates of both the bridge height and width suggests that their ratio is approximately constant, which is reflected in Fig.~\ref{fig:contact}.
There is a slight tendency of a flattening of the bridge in the initial stage of coalescence, but after approximately 1 ms, 
the bridge reaches a constant
profile. The ratio of height and width corresponds to an arc-shaped cross section with a contact angle close to the equilibrium value at both sides. During the first 1 ms, the bridge appears to be 25 \% flatter, but this deviation is at the limits of experimental resolution.

\section{Discussion}

The scaling characteristics of the coalescence of isotropic droplets in smectic liquid-crystal films can be excellently described with the lubrication model of Hack et al.~\cite{Hack2019}. The early bridge expansion is well described by by considering only the flow in a cross section along the connecting axis of the two liquid lenses.
The predictions on self-similarity are
excellently confirmed in our experiments. The self-similar solution applies over a long time range, until the bridge height reaches $\approx 75$\% of its final value. Our system allows to vary the contact angle over one order of magnitude by adjusting different temperatures in the vicinity of the clearing point of the material. 

While the scaling properties are in perfect agreement with the predictions, it turns out that the quantitative coefficients are off by one order of magnitude. Considering the known material parameters of the smectic material, surface tension and shear viscosity, the actual coalescence speed is at least an order of magnitude too slow. It is unlikely that this discrepancy arises from the uncertainties of the experiment. Rather, a systematic origin of the delay of coalescence has to be sought.
Several possible explanations may be considered. The most probable one is that the {\em effective} surface tension during coalescence is actually much smaller than assumed. The value given above is the static surface tension, which is relevant for processes that take place on time scales of several milliseconds and slower.
On such long time scales, thin freely suspended films can decrease their surface area, e. g., by creating additional smectic layers \cite{May2012,May2014}. On the other hand, expansions of the surface can be achieved if holes are torn into the upper layers of the smectic film \cite{Harth2014}, and the layers rearrange afterwards. 
When a freely suspended smectic film or a similarly perfectly ordered smectic sample undergoes such quick reorganizations of layers, it requires additional energy to create, displace or remove dislocations in the layer structure. This energy has to be delivered
by the surface energy reduction.
In the extreme of very fast processes, the smectic material keeps its surface constant, i. e. it behaves as if it had a zero effective surface tension. This is manifested, for example, by the formation of extrusions or wrinkles in laterally compressed smectic films~\cite{May2012,Harth2019}.

\begin{figure}[ht!]
	\centering
	\includegraphics[width=0.66\columnwidth]{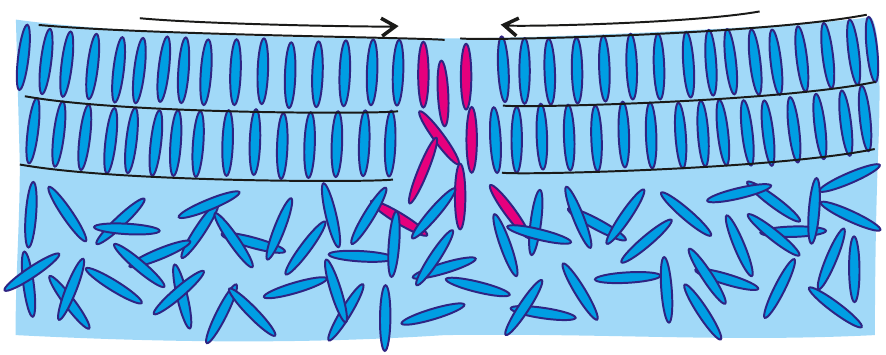} 
	\caption{{Sketch of the smectic layers at the surface of the isotropic droplet. A quick reduction of the surface cannot be compensated by lateral displacement of the smectic material. It  
	must either be compensated by melting of the surface material (red) into the isotropic phase, or by the formation of new smectic layers.}}
	\label{fig:toplayer}
\end{figure}

In the present coalescence experiment, the smectic layers on the surface of the isotropic droplets must be disposed when the surface area is reduced. Since the coalescence proceeds in the millisecond and sub-millisecond time range,
one can expect that the effective surface tension is indeed decreased respective to the stationary value. Figure \ref{fig:toplayer} sketches this scenario. A back-of-the-envelope calculation provides an estimate of the energy needed to dispose the smectic surface layers. The melting enthalpy of liquid crystals from the nematic to the isotropic phase \cite{Crucenau2008} is of the order 
of 6 kJ/kg. The energy per film area required to melt a single smectic layer at the surface (thickness approximately 3 nm) can be estimated as 0.018 N/m, this is of the order of the static surface tension itself. If the dynamic surface tension is reduced by this process, it may reach an effective 
value that is much smaller than the static surface
tension. A direct measurement of dynamic surface tensions of smectics respective to air has not been reported so far, thus it is difficult to evaluate this explanation quantitatively.

Another effect, although less important, may be the influence of the surrounding air viscosity. In fact, the bridges have submicrometer heights at least during the initial stage of coalescence. Then, one has to take into account that an air layer (viscosity $\eta_{\rm air}\approx 2\cdot 10^{-5}~\rm Pa\,s$) above and below the droplet is carried with the fluid. The dimension of this air layer is roughly given by the Saffman length
\cite{Saffman1975,Saffman1976}, $\ell_s=\eta h /(2\eta_{\rm air})\approx 350~\rm \mu m$.
Saffman's calculations apply to the mobility of a circular object moving in a thin film that has a higher viscosity than the surrounding fluid. The adaption of this model to the in-plane shear flow around the coalescing droplets is not straightforward, thus a straightforward quantitative estimation of the influence of air advection cannot be made, but it seems that this effect should be of little importance. The thickness of bridges is indeed two orders of magnitude lower than in Hack's study \cite{Hack2019}, but in their work, where the influence of the surrounding fluid was assumed to be negligible, the subphase was water with a much higher viscosity (factor 100) than air. For coalescing spherical drops~\cite{paulsen1}, it was previously shown that the outer fluid's viscosity has little influence on the early coalesence dynamics, even if it is substantially larger than the droplets' viscosity.

\section{Summary and outlook}
Summarizing, isotropic droplets in thin freely suspended smectic films provide an excellent system to measure the coalescence of flat lens-shaped droplets. The droplet coalescence is driven by the surface tension of the droplets, in contrast to flat islands in smectic free-standing films \cite{shuravin1,NguyenPhD,Stannarius2017} where the
coalescence is driven by the line tension around the islands. The scaling behaviour of the bridge shapes is excellently captured by the thin sheet model of Hack et al.~\cite{Hack2019}, but the bridge expansion proceeds an order of magnitude slower than predicted. The static surface tension of the material seems to be 
inappropriate for setting the time scale of the coalescence process, since the isotropic droplets are wetted by thin smectic surface layers whose dynamic surface tension is much smaller than the equilibrium value. It only enters the model implicitly through the contact angle, which is a static or very slowly varying variable.  

We propose to perform similar experiments in free-standing smectic films using droplets of other immiscible liquids that are not wetted by the smectic material, in order to confirm or discard our hypothesis. The problem is to find an immiscible material with a surface tension that is only slightly larger than that of the smectic. Otherwise, the contact angle will be too large and the interference rings are much denser so that they cannot be distinguished unambiguously. This will render the height profile measurement impracticable. Furthermore, the lubrication approximation may no longer be applicable. Theoretical studies accounting for a more complex surface rheology or incorporating the dynamics of smectic layer melting/reorganization may shed light on the mechanism of the reduction of the effective surface tension in our coalescing molten droplets covered by few smectic layers.

\section*{Acknowledgments}
The  authors  cordially thank  Alexey Eremin, Kirsten Harth and Michiel Hack for stimulating and fruitfull discussions. The authors also acknowledge the German Science Foundation (DFG) for support within project STA 425/40-1, the German Aerospace Center (DLR) for support within project 50WM1744. C. K. acknowledges support by a Landesstipendium Sachsen-Anhalt.

\providecommand*{\mcitethebibliography}{\thebibliography}
\csname @ifundefined\endcsname{endmcitethebibliography}
{\let\endmcitethebibliography\endthebibliography}{}

\end{document}